# A Note on the Membrane Computer

Ammar Adl

(ammaradl@gmail.com)

Amr Badr

(a.badr.fci@gmail.com)

Ibrahim Farag

(i.farag@gmail.com)

Computer Science Department
Faculty of Computers and information
Cairo University

## Abstract

*Inspired by the emergent membrane computing (P Systems) concepts, some efforts are carried out introducing simulation models, some are software oriented, and others are hardware, yet all are applied with the current vision of the conventional computers, based on "Von Neumann architecture", which is a sequential design in its essence. We think that these models will need –as a consequent result– to a new architecture exposing a true parallel design, in this paper; we try to investigate and introduce a global view for how it would be like to have such architecture. The main goal is to point out to this direction broadly, suggesting that it might be useful considering some aspects, like the need for a new definition of an operating system and its programs, which will eventually lead to a higher scope: the membrane computer.*

## 1. Introduction

Contemporary computers have increased their performance since the early days of computing, but this trend is limited by physical laws. Although many real-life problems can be solved in reasonable time, other relevant problems need an exponential amount of resources (time or space) to be solved. Simulating P Systems is based on the current computational processing concepts; some problems are permanently arising, like communication bottlenecks, limited processors numbers, global system clock, synchronization, and others. Several authors have implemented parallel simulators for transition P systems on clusters [3], and many others have designed P systems hardware implementations. [8]

Here, a different look could be set upon the concept of membrane computing device itself, in an implementation sense. If we can represent an operating system with a membrane model, this might lead us to explore the idea of whole new computer architecture, which will consequently points us to some new concepts of how programs will be developed, and what aspects should they depend upon.

In this work, we will try to reach out beyond what is considered to be a "Simulation". Eventually there will be a need for a proposed architecture to the upcoming era of membrane computing, so first, we will consider some current models, efforts, and then try to evolve with the results, and cross that barrier to a further look, in pursuit of the (Membrane Computer).

## 2. Preliminaries

Recalling definition from [8]; P systems consist of a set of *syntactic* components: a celllike *membrane structure* (it is a hierarchical rooted tree of compartments that delimit *regions*, where the root is called *skin*), *multiset of objects* (corresponding to chemical substances present inside the compartments (membranes) of a cell), and *rules* (corresponding to chemical reactions that can take place inside the cell) executed in a synchronous nondeterministic maximally parallel manner.

The *semantics* of P systems are defined through a nondeterministic synchronous model. A computation of a P system is made by steps called *configurations* (which identifies an instantaneous state of the system: the family of multisets and the membrane structure), assuming a global clock that synchronizes the execution.

The computation starts always with an *initial configuration* of the system, where the input data of a problem is encoded. [6]
The *transition* from one configuration to the next is performed by applying rules to the objects placed inside the regions. A computation of the system is a tree of configurations, which is made by transitions until reaching a *halting configuration*, where no more rules can be applied. The result of a halting computation is usually defined through the multiset associated with a specific output membrane, or the *environment*, which is the space out of the skin. [4]
In the P systems with active membranes model, every elementary membrane is able to divide itself by reproducing its content into a new membrane. Specifically, this model is a construct of the form:

$$\Pi = (O, H, \mu, \omega_1, ..., \omega_m, R)$$

Where $m \geq 1$ is the initial degree of the system; O is the alphabet of *objects*, H is a finite set of *labels* for membranes; μ is a membrane structure (a rooted tree), consisting of m membranes injectively labeled with elements of $H, \mu, \omega_1, ..., \omega_m$ are strings over O, describing the *multisets of objects* placed in the m regions of μ; and R is a finite set of *rules*, where each rule is one of the following types: *Evolution rules*, *Division rules*, *Dissolution rules* and *Communication rules* (*Send-out* or *Send-in*).

P systems can be used for addressing the efficient resolution of decision problems. This kind of problems requires either a *yes* or *no* answer [3]. In this sense, we consider *recognizer P systems* as P systems with external output (the results of halting computations are encoded in the environment) such that the *yes* or *no* answer is sent in a halting configuration.

## 3. Hardware simulation

There were some efforts in the perspective of hardware simulation like Parallel Computing on GPU (graphical processing unit) In this section, we will recall some efforts in [10], [8], [1], giving a brief introduction to work done on the NVIDI, a Tesla C1060 GPU is provided, and also the CUDA parallel programming model presented in [2][1]. Each thread has its own thread execution state and can execute an independent code path. The SMs execute threads in a Single-Instruction Multiple-Thread (SIMT) fashion (see [2] [1]). In the CUDA parallel programming [13], an application consists of a sequential code (host code) that may execute parallel programs known as kernels on a parallel device. The programmer organizes the threads into a grid of thread blocks. A thread block in CUDA is a set of threads that execute the same program (kernel) and cooperate to obtain a result through barrier synchronization and a per-block shared memory space, private to that block.

The selection stage consists of the search for the rules to be executed in each membrane; the execution stage consists of the execution of the rules previously selected. It presents a need of global synchronization among each phase. The input data for the selection stage consists of the description of the membranes with their multisets (strings over the working alphabet O, labels associated with the membrane in H, etc...); the output data of this stage is the set of selected rules per membrane.

The execution stage applies the rules previously selected on the selection stage, and the membranes can vary including new objects, dissolving membranes, dividing membranes, etc, obtaining a new configuration of the simulated P system. It is an iterative process until a halting configuration is reached. The execution of those rules were designed as different CUDA kernels, one kernel per each kind of rule (send-in communication, send-out communication, dissolution and division), giving a result of the execution stage.

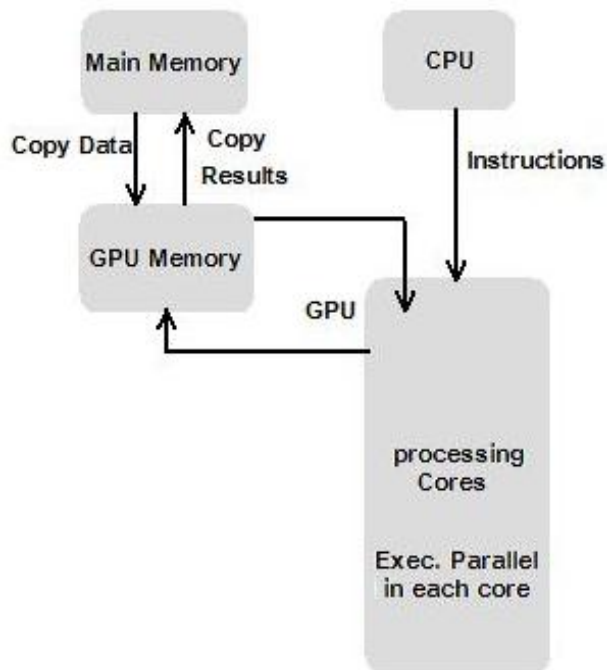

**Figure 1. CUDA system.**

## 4. Software Simulation

If one would like to simulate the function of an operating system running on a mainstream (one-CPU) computer, it would be necessary to build into the membrane system a synchronizing system like the following one: [1]

$$\Pi = (V, T, C, H, \mu, \omega_1, \omega_2, (R_1, \rho_1), (R_2, \rho_2))$$

In this example the membrane system produces no output. Membrane 1 is running like an operating system which controls membrane 2. The clock signal splits between the two membranes, only one of them is operating at each step.

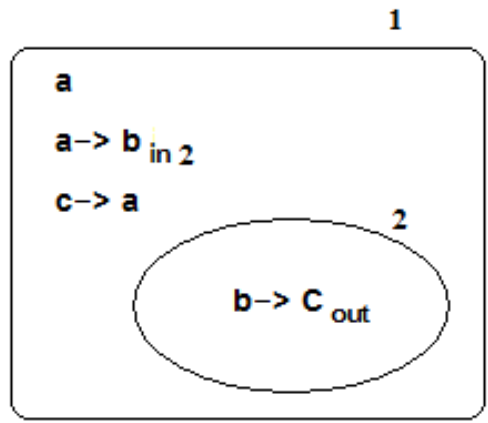

**Figure 2. Synchronization in a P- System.** [1]

$$\Pi = (V, T, C, H, \mu, \omega_1, \omega_2, \omega_3, \omega_4, (R_1, \rho_1), ..., (R_4, \rho_4))$$
*Where*
$V = \{a_4\} \cup V'$;
$\mu = [_1[_2[_3]_3]_2[_4]_4]_1$;
$\omega_1, R_1 = \Gamma_1 \cup \{a_4 \to (a_4)_{in4}\}$,
$\omega_2, R_2 = \Gamma_2 \cup \{a_4 \to (a_4)_{out}\}$,
$\omega_3, R_3 = \Gamma_3 \cup \{a_4 \to (a_4)_{out}\}$,
$\omega_4, R_4 = \Gamma_4$.

Where $a_4$ can occur only on right-hand-side of rules of $(\Gamma_1 \cup \Gamma_2 \cup \Gamma_3)$. [1]

In this example $a_4$ is a special object; only region 4 can deal with it other regions may produce and must transfer it to the direction of the target (region4). If $\Gamma_3$ contains rules like $a \to aa; a^n \to a4$ where n is an arbitrary fixed integer $(n \geq 2)$, and $\Gamma_4$ contains a rule like $a_4 a_4 \to a_4$ then this example will be just like the classical producer-consumer problem. From time to time membrane 3 can produce objects $a_4$ and sends them to membrane 4 which reduce their number.

## 5. Membrane System as an OS

If we would like to have an efficient simulation of a membrane computation with a general operating system (with a mainstream hardware), we would have serious restrictions: we would have to use a fixed number of membranes (processes), because of the fixed number of CPUs, deterministic computation with random value generation can simulate several language generating P-systems.

Several operating systems have versions using several (i.e. up to 8, 32, and 128) processors, using active membranes with (theoretically) not bounded number of regions an effective simulation needs a dynamic system which can increase the number of processors. In these systems the resources, such as the number of computers (processors) can dynamically be changed according to the needs of the computation. There are special types of programming techniques (that play similar role as a network operating system) required by GRIDs. GRID techniques are used in several computations in various fields, where the same type of computation is needed for several cases. The computation with membranes has the same phenomenon: the computational processes in the deepest levels are highly independent, and as a general mapping like in [1]:

| Membrane System | Operating System |
|---|---|
| Skin Membrane | OS Shell |
| Membrane | Program |
| Rules | Program instructions |
| Objects | Program Data |
| Clock | Clock |
| Tree Structure of Membranes | Tree Structure of Processes |

**Figure 3. General MP – OS mapping [1]**

From the above we believe that there will be a need for a general architecture to formulate this environment of membrane computing, the conventional and already existing models are suited for simulation only, the true power of membrane computing will not be fully addressed unless architecture is introduced to empower full utilization of this processing model.

As for the current computers, the *von Neumann architecture* which was a design model for a stored-program, digital computers that uses a processing unit and a single separate storage structure to hold both instructions and data. Such computers implement a Universal Turing machine and have a sequential architecture.

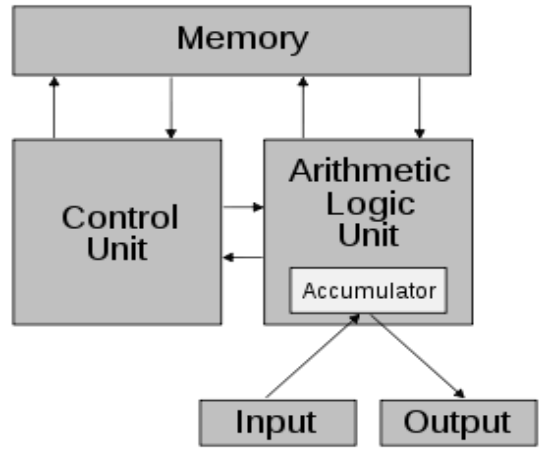

**Figure 4. Design of the von Neumann architecture.**

The proposed architecture design will have to overcome the global clock issue, and processing elements must be reconsidered from the beginning.

A new look for many constituents should be pursued, for instance:

Given the ability of the cell to produce membranes and carry out many processes at the same time, we think that the cell itself will be representing a specialized operating system (Membrane OS) – MOS for short, as we have seen from the simulation above the outer membrane itself could be considered the OS shield, this OS is not a general purpose one, yet it gives a container to small programs carrying out some instructions, those programs are the inner membranes themselves, this will lead to newer concepts of programming pieces of logic, writing (Membrane Programs) with some aspects like (Membrane Oriented Programming).

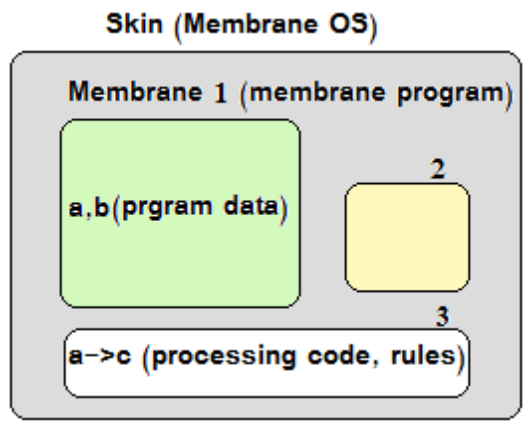

**Figure 5. Illustrating the concept of MOS**

Every MOS will be permitting some connection points to accept initial configuration and incoming data, then it kicks off a chain of commands to its programs (membranes) for processing the data, all this could be done with a local clock into the MOS itself, apart from other MOSs.

With a tissue of micro MOSs, data could be processed in true parallelism, given that view, this could lead to something like self reproductive small devices of computation, which at some point of time, If failure occurs, the MOS by itself can render a second version of itself with the same initial configuration copying all initial membranes (programs) to the newer one. This is through a reproduction end, which carries rules of reproducing another copy of the MOS, with its initial configuration.

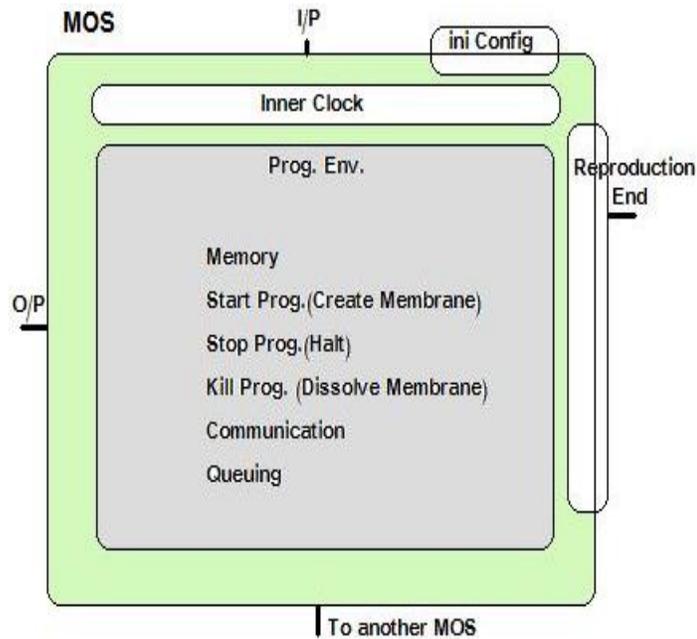

**Figure 6. Illustration for MOS.**

Apparently this thinking approach would imply a global view for a whole computing architecture, which could engage some bio-concepts of resources management, true parallelism, clock localization, elimination of communication bottlenecks, and other existing problems.

Here we can imagine that the abstract view for such architecture would be like the following: The outer layer represents the output and input segments, the core control unit will be in charge of global tasks, and control actions, the communication cord is the system's global bus - not like the usual bus used now - just a channel to maintain all systems components in a linked state, used to carry major actions, system monitoring, and finally some groups of different MOSs, "colonies-like" MOSs put up together, each group for a specific type of tasks, the tissue provides the suitable resources for the processing environment.

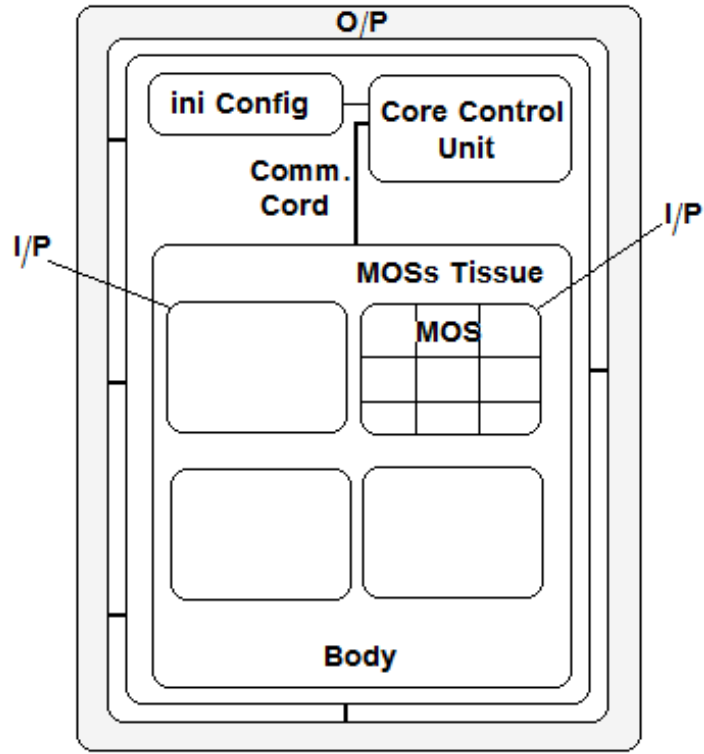

**Figure 7. Membrane Computer Architecture**

## 6. Conclusion

Approaching a computational architecture that mimicking the living cell in its processing model and extending such features to a full computational schema will eventually lead to a global architecture for the (Membrane Computer). With such way of thinking some points will arise, as a consequence of the membrane computer:

Membrane programming language (MPL) not meant a simulation one like the (pLingua) [7], but for true membrane programs. Also there will be a need for membrane virtual machine (MVM), all this will lead to new programming concepts like stated above (MOP) - Membrane Oriented Programming. All these points with more investigation on the core abstract architecture and its major components will shape the future of this research.

# References

[1]   Benedek Nagy, Laszlo Szegedi, Membrane Computing and Graphical Operating Systems, Journal of Universal Computer Science, vol. 12, no. 9 (2006)

[2]  E. Lindholm, J. Nickolls, S. Oberman, J. Montrym. NVIDIA Tesla: A unified graphics and computing architecture. IEEE Micro, 28, 2 (2008).

[3]  G. Ciobanu, G. Wenyuan, C. Martin-Vide, G. Mauri, G. Paun, G. Rozenberg, A. Salomaa (eds.). P systems running on a cluster of Computers. Workshop on Membrane Computing, vol. 2933 (2004).

[4]  G. Ciobanu, M.J. Perez–Jimenez, G. Paun, Applications of membrane computing. Natural Computing Series, Springer, (2006).

[5]  G. Paun. Computing with membranes. Journal of Computer and System Sciences, 61, 1 (2000), and Turku Center for Computer Science-TUCS Report No 208.

[6]  G. Paun. Membrane Computing: An Introduction. Springer, (2002).

[7]   J. D. Owens, D. Luebke, N.Govindaraju, M. Harris, J.Krger, A.E. Lefohn, T.J. Purcell. A survey of general–purpose computation on graphics hardware. Computer Graphics Forum, 26, 1 (2007).

[8]  Jose M. Cecilia, Gines D. Guerrero, Jose M. Garcia, Miguel A. Martinez–del–Amor, Ignacio Perez–Hurtado, Mario J. Perez– Jimenez. A massively parallel framework using P systems and GPUs, Symposium on Application Accelerators in High Performance Computing, July 2009.

[9]   L. Fernandez, V.J. Martinez, F. Arroyo, L.F. Mingo. A hardware circuit for selecting active rules in transition P systems. Proceedings of the Seventh International Symposium on Symbolic and Numeric Algorithms for Scientific Computing (2005).

[10]  M. A. Martinez-del-Amor, I. Perez-Rtado, M.J. Perez-Jimenez, J.M. Cecilia, G.D. Guerrero, J.M. Garcia. Simulating Active Membrane Systems Using GPUs (2009)

[11]  M. Garcia-Quismondo, R. Gutierrez- Escudero, M.A. Martinez-del-Amor, E. Orejuela-Pinedo, I. Perez-Hurtado. PLingua 2.0: A software framework for cell–like P systems. Int. J. of Computers, Communications and Control, Vol. IV (2009).

[12]  S. Ryoo, C. Rodrigues, S. Baghsorkhi, S. Stone, D. Kirk, W. mei Hwu. Optimization principles and application performance evaluation of a multithreaded GPU using CUDA. Proceedings of the 13th ACM SIGPLAN Symposium on Principles and Practice of Parallel Programming, (2008).

[13]  S. Ryoo, C. Rodrigues, S.S. Stone, J.A. Stratton, Sain-Zee Ueng, S.S. Baghsorkhi, W.W. Hwu. Program optimization carving for GPU computing. J. Parallel Distrib. Comput., 68, 10 (2008).

[14] V. Nguyen, D. Kearney, G. Gioiosa. An algorithm for non-deterministic object distribution in P systems and its implementation in hardware, Membrane Computing, 9th International Workshop (2009).